\documentclass{emulateapj}
\usepackage{apjfonts}

\lefthead{JUNG ET AL.}
\righthead{REEVALUATING EARTH-MASS DETECTION PROBABILITY}

\begin{document}
\title{REEVALUATING OF THE FEASIBILITY OF GROUND-BASED 
EARTH-MASS MICROLENSING PLANET DETECTIONS}

\author{
Youn Kil Jung$^{1}$,
Hyuk Park$^{1}$,
Cheongho Han$^{1,}\footnotemark[2]$,
Kyu-Ha Hwang$^{1}$,
In-Gu Shin$^{1}$ and
Joon-Young Choi$^{1}$
}

\footnotetext[2]{Corresponding author}

\bigskip\bigskip
\affil{$^{1}$Department of Physics, Institute for Astrophysics, 
  Chungbuk National University, Cheongju 371-763, Korea}

\begin{abstract}
One important strength of the microlensing method in detecting 
extrasolar planets is its high sensitivity to low-mass planets. 
However, it is often believed that microlensing detections of 
Earth-mass planets from ground-based observation would be difficult 
due to the limit set by finite-source effects.  This view comes 
from the previous estimation of the planet detection probability 
based on the fractional deviation of planetary signals, but proper 
probability estimation requires to additionally consider the source 
brightness, which is directly related to the photometric precision. 
In this paper, we reevaluate the feasibility of low-mass planet 
detections considering photometric precision for different populations 
of source stars.  From this, it is found that contribution of the 
improved photometric precision to the planetary signal of a giant-source 
event is big enough to compensate the decrease of the magnification 
excess caused by finite-source effects. As a result, we find that 
giant-source events are suitable targets for Earth-mass planet 
detections with significantly higher detection probability than 
events involved with source stars of smaller radii and predict that 
Earth-mass planets would be detected by prospective high-cadence 
surveys.  
\end{abstract}


\section{Introduction}
Searches for extrasolar planets by using the microlensing method is 
being conducted toward the Galactic bulge field (OGLE: \citet{udalski03}, 
MOA: \citet{bond01}, \citet{sumi03}, Wise: \citet{shvartzvald12}). Due to the
high sensitivity to planets that are difficult to be detected by other methods, 
the method is important for the comprehensive understanding of the formation 
and evolution of planets in various types of stars \citep{mao91,gould92,gaudi12}.

One of the important strength of the microlensing method is that it is sensitive to 
low-mass planets. This is because the amplitude of a microlensing planetary signal 
does not depend on the planet mass for a point source, 
although the duration of the signal becomes shorter with the decrease of the planet mass. 
In practice, the low mass limit of a microlensing planet is set by 
finite-source effects which wash out planetary signals. 
For giant source stars, the size of the caustic induced by an Earth-mass planet is 
equivalent to the angular size of the source star and 
thus the planetary signal is significantly weakened 
due to severe finite-source effects. 
For events associated with a main-sequence (MS) star, 
on the other hand, the attenuation of the planetary signal is mild 
but poor photometric precision caused by the source faintness 
limits secure detections from ground-based observation. 
As a result, it is often believed that detecting Earth-mass planets 
from ground-based observation would be difficult.

The difficulty of detecting Earth-mass planets from ground-based microlensing 
observation was first pointed out by \citet{bennett96}. 
The basis of their result lies on their estimation of the planet detection probability 
based on the simulation of planetary lensing events involved with 
various types of source stars. In their simulation, they computed 
the fractional deviation of the lensing magnification, $(A-A_{0})/A_{0}$, 
and estimated the detection probability by imposing a threshold deviation. 
Here $A$ and $A_{0}$ represent the lensing magnifications with and without 
the presence of the planet, respectively. With this criterion, 
the planet detectability is mostly decided by the severeness of finite-source effects 
and thus they reached a conclusion that the detection probability of 
giant-source events would be significantly lower than the probability of 
events associated with faint source stars with smaller radii. 
Based on this result, \citet{bennett02} proposed a space-based microlensing experiment 
in search for Earth-mass planets by resolving faint MS stars.

However, proper estimation of the planet detection probability requires to 
additionally consider the source brightness. 
This is because the strength of a planetary signal 
$\Delta\chi^2=\sum_{i}{(F_{i}-F_{0,i})}^2/{\sigma_{i}}^2$ depends not only on 
the amplitude of the planetary deviation, $F-F_{0}$, 
but also on the photometric uncertainty, $\sigma$, 
which is directly related to the source brightness. 
Here $F$ and $F_{0}$ represent the observed source fluxes 
with and without the planet, respectively. 
In addition to the direct decrease of photon noise with 
the increased photon count, photometric precision further 
depends on the source brightness because bright source events 
are likely to be less affected by blending. 

In this paper, we reevaluate the feasibility of ground-based 
detections of Earth-mass planets by additionally considering 
the dependence of photometric precision and blending on the source type. 
In Section 2, we describe the simulation of planetary lensing events 
conducted for the estimation of the probability. 
In Section 3, we present results from the analysis. 
We summarize the result and conclude in Section 4.

\section{Simulation}
In order to evaluate the feasibility of ground-based Earth-mass 
planet detections, we conduct simulations of Galactic microlensing events. 
The simulation is based on representative lensing events following 
current and/or prospective Galactic microlensing surveys.

For the lens, we assume that the mass of the primary lens is 
$0.3$ $M_{\odot}$ by adopting that of the most common lens 
population of low-mass stars \citep{han95}. 
Then, the mass ratio of an Earth-mass planet to the lens 
is $q \sim 10^{-5}$. We adopt an Einstein time scale of $t_{\rm E} = 20$ days 
and assume that events are observed with a 10 minute cadence 
following the observational strategy of the prospective ground-based 
survey of the Korean Microlensing Telescope Network 
\citep[KMTNet:][]{kim10}. Following the OGLE lensing survey, 
we assume that images are taken in $I$ band.

\begin{deluxetable}{lcccc}
\tablewidth{0pt}
\tablecaption{Source Characteristics\label{table:one}}
\tablehead{
\colhead{source} & 
\colhead{$I_{\rm 0}$} & 
\colhead{$F_{\rm b}/F_{\rm s}$} & 
\colhead{$\rho_*$} &
\colhead{$\Gamma_I$}
} 
\startdata
main-sequence & 19.5 & 4.00 & 0.001 & 0.40 \\
subgiant      & 18.0 & 1.00 & 0.003 & 0.45 \\
giant         & 16.5 & 0.25 & 0.010 & 0.48 
\enddata
\vspace{0.05cm}
\tablecomments{$I_{\rm 0}$: unlensed, unblended source brightness,
$F_{\rm b}/F_{\rm s}$: blend to source flux ratio,
$\rho_* = \theta_*/\theta_{\rm E}$: source radius normalized 
by the Einstein radius,
$\Gamma_I$: linear limb-darkening coefficient
}
\end{deluxetable}

For the source, we test three representative stellar populations of Galactic bulge stars 
including MS, subgiant, and giant stars. 
For the source radii $\theta_*$ normalized by the Einstein radius $\theta_{\rm E}$, 
we adopt $\rho_* = \theta_*/\theta_{\rm E} = 0.001$, $0.003$, and $0.010$ 
for the individual source stars, which roughly correspond to 
the physical source radii of the $1$ $R_{\odot}$, $3$ $R_{\odot}$, and $10$ $R_{\odot}$, 
respectively. Considering the stellar types and distance to the Galactic bulge, 
and adopting an average extinction $A_I = 1.0$ \citep{nataf13}, 
we assume that the unlensed, unblended, apparent $I$-band magnitudes of 
the individual source stars are $I_{\rm 0} = 19.5$, $18.0$, and $16.5$, 
respectively. In Table \ref{table:one}, 
we summarize the characteristics of the individual source stars.

\begin{figure}[ht]
\epsscale{1.15}
\plotone{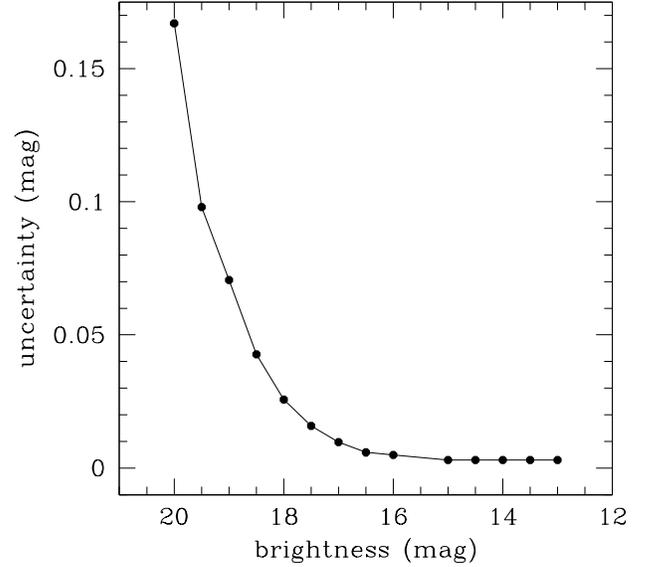}
\caption{\label{fig:one}
Adopted relation between the photometric uncertainty and 
the source brightness. The relation is obtained based on actual lensing events 
observed by the OGLE survey.
}\end{figure}

For realistic simulations of lensing light curves, 
photometric errors are estimated by using the relation 
between the photometric uncertainty and the source brightness 
that is obtained based on actual events observed by 
the OGLE survey. 
For this, we choose multiple number of events with wide spans of 
lensing magnification and take an average value 
as a representative uncertainty for a given source brightness. 
We note that systematics of photometry is taken into consideration 
because the adopted magnitude-error relation is estimated based 
not on theoretical assumptions of systematics 
but on actual data resulting from systematics. 
Figure \ref{fig:one} shows the magnitude-uncertainty relation. 
At faint magnitudes, photometry is limited by photon count, 
resulting in rapid decrease of error as a star becomes brighter. 
At very bright magnitudes, on the other hand, photometry is mostly dominated by 
non-photon noise, e.g. read-out noise, dark current, etc, 
and thus photometry does not improve for $I \leq 14$. 
In the region $I \geq 16$, it is found that the uncertainty decreases 
very rapidly as the source becomes brighter. 
For example, sub-milli magnitude level photometry is possible for a giant star, 
while the uncertainty for a MS star is $\sigma \sim 0.1$ mag. 
We assume that photometry follows Gaussian distribution. 

Photometric precision in lensing observation depends 
not only on the source brightness but also on the blended light 
from unlensed neighboring stars. Considering that faint stars are 
more affected by blending, we adopt blend to source flux ratios of 
$F_{\rm b}/F_{\rm s} = 4.0$, $1.0$, and $0.25$ 
for the individual source populations. 
With these ratios, the source brightness of a lensing event varies as
\begin{equation}
I(t) = I_{\rm base} - {\Delta}I(t),
\label{eq1}
\end{equation}
where $I_{\rm base} = I_{\rm 0} - 2.5{\rm log}(1+F_{\rm b}/F_{\rm s})$ 
is the blended baseline source brightness, 
${\Delta}I(t) = 2.5{\rm log}A_{\rm obs}(t)$ is 
the apparent brightening of the source, 
and $A_{\rm obs}(t)=[A(t)+F_{\rm b}/F_{\rm s}]/(1+F_{\rm b}/F_{\rm s})$ 
is the apparent lensing magnification \citep{han99}.

To produce light curves affected by finite-source effects, 
we use the ray-shooting method \citep{kayser86,schneider87}. 
In this method, one shoots uniform rays from the observer to the lens plane, 
calculates the angle of deflection due to the individual lens components, 
and collects rays in the source plane. The deflection angle is computed by the 
lens equation
\begin{equation}
\zeta=z-{\epsilon_{1}\over \overline{z}-\overline{z}_{\rm L,1}}
-{\epsilon_{2}\over \overline{z}-\overline{z}_{\rm L,2}},
\label{eq2}
\end{equation}
where $\epsilon_{i}$ is the mass fraction of each lens component, 
$\zeta$, $z_{{\rm L},i}$, and $z$ represent the complex notations of 
the source, lens, and image positions, respectively, 
and $\overline{z}$ denotes the complex conjugate of $z$ \citep{witt90}. 
Then, finite-source magnifications 
are computed as the ratio of the number density of rays on 
the surface of the source star to the density on the observer plane. 
In computing finite-source magnifications, we additionally consider 
the variation of the source surface brightness caused by 
limb-darkening effects. For this, we model the surface brightness profile as 
\begin{equation}
S_{I}\propto 1-\Gamma_{I} \left(1-{3\over 2}\cos \phi \right),
\label{eq3}
\end{equation}
where $\Gamma_{I}$ is the linear limb-darkening coefficient and $\phi$ 
is the angle between the line of sight toward the source star and 
the normal to the source surface \citep{albrow01}. 
The adopted values of the limb-darkening coefficients are 
listed in Table \ref{table:one}.

\begin{figure*}[ht]
\epsscale{0.85}
\plotone{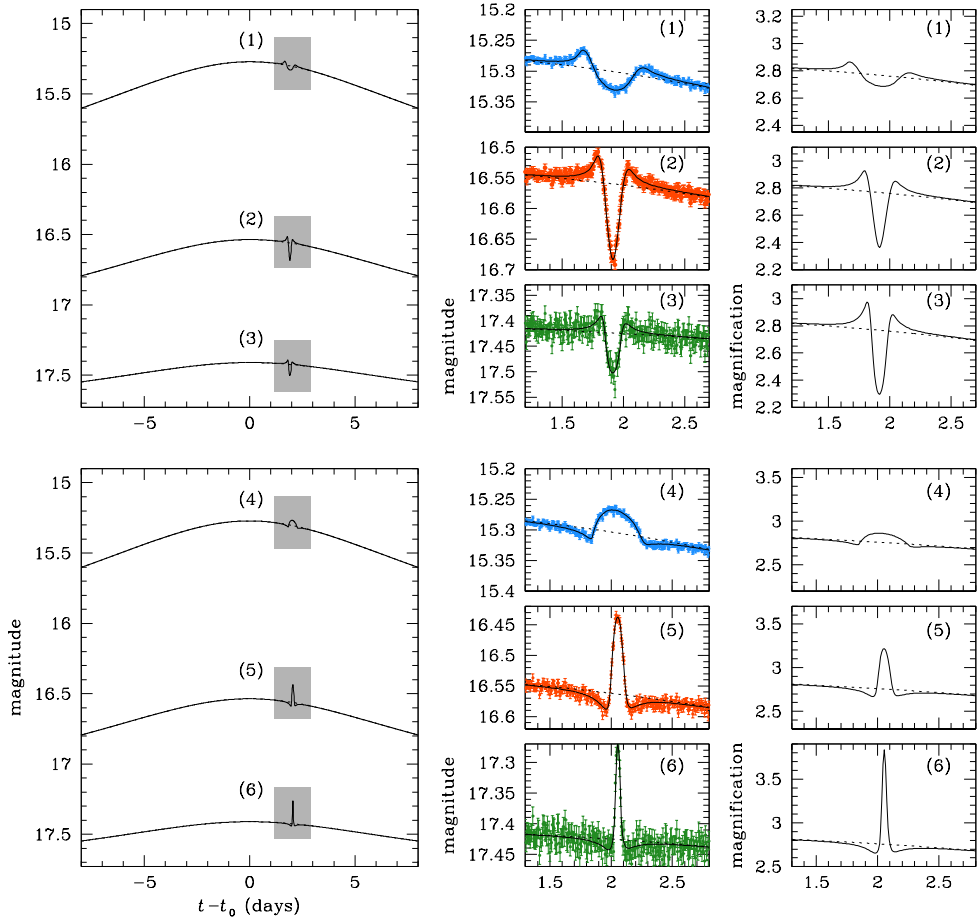}
\caption{\label{fig:two}
Example light curves of microlensing events produced by simulation. 
The upper and lower sets of panels show light curves produced by 
planets with projected separations $s = 0.83$ and $1.20$, respectively. 
The three light curves in each set of panels represent those of events 
occurred on giant (top), subgiant (middle) and 
MS (bottom) source stars. 
The shaded region on each light curve represents the planetary 
perturbation region and the enlarged view is presented 
in the middle and right panels 
marked by the corresponding number. The dotted curve for each light curve 
represents the light curve without the planet. 
}
\end{figure*}

In Figure \ref{fig:two}, we present example lensing light curves 
produced by simulation. The upper and lower sets of panels 
show the two typical cases of planetary perturbations 
produced by close (upper panels) and wide (lower panels) 
planets, respectively. Here the terms ``close'' and ``wide'' denote the cases 
where the projected planet-star separation is smaller 
and bigger than the Einstein radius, respectively. 
For each set of panels, we present three light curves associated with 
giant (top), subgiant (middle), and MS (bottom) source stars. 
We note that the lensing magnifications of the individual light curves 
are identical but the light curves appear different 
due to the differences in the source brightness and the blend to source flux ratio. 
The three small middle panels show the enlarged view of the 
perturbation regions (shaded region in the left panel) presented in ``magnitude''. 
Also presented in the right panels are the light curves of the perturbations 
in terms of the lensing ``magnification''.

\section{Detection Probability}
With light curves produced from simulation, 
we then estimate the planet detection probability. 
The probability is estimated as the fraction of events 
with noticeable planetary signals among all tested events 
under the assumption that source trajectories 
with respect to the star-planet axis are randomly oriented.

Estimating the detection probability requires to produce 
a large number of finite-source light curves which demand heavy computation. 
For efficient estimation of the probability, 
we use the map-making method \citep{dong06}. 
In this method, a pixel map of rays covering a large area 
on the source plane is stored in the buffer memory of a computer and 
light curves are produced from rays in the pixels of the map 
located along the source trajectory. This method is useful 
in producing many light curves resulting from different source trajectories 
without additionally shooting rays for each light curve.

In our estimation of the probability, we consider events 
with impact parameters (normalized by $\theta_{\rm E}$) 
of the lens-source approach $u_0 < 1 $, i.e. events produced by 
the source approach within the Einstein ring of the primary lens.
Microlensing planets can be detected for events with $u_0 > 1$ 
where the planetary signal stands on the weak or 
even no base lensing magnification of the primary 
\citep{han05, sumi11, bennett12}, 
but we do not consider these cases in this work.

\begin{figure}[ht]
\epsscale{1.15}
\plotone{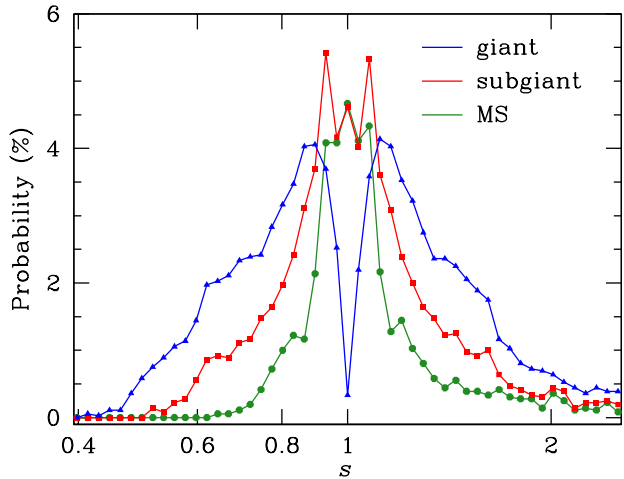}
\caption{\label{fig:three}
The probability of detecting Earth-mass planets 
as a function of the projected planetary separation 
for different populations of source stars. 
The separation is normalized by the Einstein radius.
}
\end{figure}

We estimate the probability by setting a threshold value 
$\Delta\chi^2_{\rm th}$ as a criterion for the detectability of 
planetary signals. 
\citet{gould10} extensively discussed about the choice of 
$\Delta\chi^2_{\rm th}$ and pointed out that $\Delta\chi^2_{\rm th}$ 
should be high enough 
in order to avoid false-positive signals arising from systematics in data. 
We therefore adopt a conservative value of $\Delta\chi^2_{\rm th} = 500$. 
We note that the trends of the probability dependence on the source type 
and the planetary separation for different values of $\Delta\chi^2_{\rm th}$ 
are consistent regardless of the choice of threshold values within 
$300 \leq \Delta\chi^2_{\rm th} \leq 700$.

It is known that intrinsic source variability, especially for giant stars, 
can induce short-term variation in lensing light curves \citep{gould13}. 
However, resulting deviations, in general, 
have different forms from those induced by planets and 
thus they can be easily distinguished from follow up photometric and 
spectroscopic observation. We, therefore, do not consider 
contamination of planetary lensing signals by source variability.

In Figure \ref{fig:three}, we present the estimated detection probability 
as a function of the projected planet separation normalized by 
the Einstein radius, $s$, for three different populations of source stars. 
From the distributions, it is found that the probability is 
substantially higher 
for giant-source events than those of events involved with source stars 
of smaller radii throughout the planetary separations 
except for a narrow region around $s = 1$.\footnotemark[1] 
The result is contrary to that of \citet{bennett96} 
where they found a smaller probability for a larger source size.

The cause of the opposite results between our and Bennett \& Rhie's analyses 
is found from the comparison of 
the planetary perturbations presented in 
the brightness (magnitude) and magnification, 
which are shown in the middle and right panels of Figure \ref{fig:two}. 
As pointed out by \citet{bennett96}, 
magnification excess of a planetary deviation 
decreases with the increase of the source size 
due to finite-source effects. However, a source star 
becomes brighter as its size increases and the contribution of 
the improved photometric precision to the planetary signal is 
big enough to compensate the decrease of the magnification excess 
and to result in stronger planetary signals. 
For example, we find that the planetary signal $\Delta\chi^2 = 2878$ 
of the giant-source event marked by (1) is much stronger than 
the signal $\Delta\chi^2 = 322$ of the MS-source event marked by (3) 
despite that the former event suffers from 
much more severe finite-source effects.

\footnotetext[1]{
The dip in the probability distribution 
for planets with $s \sim 1$ is due to the weak magnification excess of 
the caustic region relative to a single lens \citep{gaudi11}. 
As a result, planetary signals are more vulnerable to finite-source effects, 
causing lower detection probability for giant-source events than 
events with smaller source radii. 
}

\section{Summary and Discussion}
In order to check the previous result indicating the difficulty in 
detecting Earth-mass planets from ground-based observation, 
we reevaluated the detection feasibility by additional 
considering the dependence of photometric precision on 
source populations. From the analysis based on 
realistic simulation of lensing events, 
we found that giant-source events were suitable targets 
for Earth-mass planet detections with substantially higher 
detection probability than events involved with source stars 
of smaller radii. We found the reason for the opposite result 
to the previous one is that the contribution of the improved 
photometric precision to the planetary signal of 
a giant-source event is big enough to compensate the decrease of 
the magnification excess caused by finite-source effects.

Although no Earth-mass planet has been detected yet,
the competing effects of extended source and source brightness 
are well illustrated by microlensing planets detected 
through the channel of high-magnification events. 
A good example is the planetary event MOA-2007-BLG-400 \citep{dong09}. 
Although the mass of the planet $(\sim 0.5 - 1.3$ $M_{\rm Jup})$ 
discovered in the event is much heavier than the Earth, the event is 
similar to a giant-source Earth-mass planetary event in the sense that 
the caustic was substantially smaller than the source and 
the planetary deviation occurred when the source was bright. 
Despite that the planetary deviation was greatly attenuated 
by severe finite-source effects, the signal was detected 
with the large significance of $\Delta\chi^2 = 1070$ and 
the planetary deviation was unambiguously ascertained. 
Considering that the detection was possible mostly thanks to 
the high photometric precision of the bright source, 
giant-source events would be suitable targets 
for the detection of low-mass planets. 
Therefore, we predict that Earth-mass planets 
would be detected from future ground-based high-cadence lensing surveys.

\acknowledgments
This work was supported by Creative Research Initiative Program
(2009-0081561) of National Research Foundation of Korea.


\begin{thebibliography}{99}

\bibitem[Albrow et al.(2001)]{albrow01}
Albrow, M. D., An, J., Beaulieu, J.-P., et al. 2001, \apj, 549, 759

\bibitem[Bennett \& Rhie(1996)]{bennett96}
Bennett, D. P., \& Rhie, S. H. 1996, \apj, 472, 660

\bibitem[Bennett \& Rhie(2002)]{bennett02}
Bennett, D. P., \& Rhie, S. H. 2002, \apj, 574, 985

\bibitem[Bennett et al.(2012)]{bennett12}
Bennett, D. P., Sumi, T., Bond, I. A., et al. 2012, \apj, 757, 119

\bibitem[Bond et al.(2001)]{bond01}
Bond, I. A., Abe, F., Dodd, R. J., et al. 2001, \mnras, 327, 868

\bibitem[Dong et al.(2006)]{dong06}
Dong, S., Depoy, D. L., Gaudi, B. S., et al.\ 2006, \apj, 642, 842

\bibitem[Dong et al.(2009)]{dong09}
Dong, S., Bond, I. A., Gould, A., et al. 2009, \apj, 698, 1826

\bibitem[Gaudi(2011)]{gaudi11}
Gaudi, B. S. 2011, Exoplanets, ed. S. Seager (Tucson, AZ: Univ. Arizona Press), 79

\bibitem[Gaudi(2012)]{gaudi12}
Gaudi, B. S. 2012, \araa, 50, 411

\bibitem[Gould \& Loeb(1992)]{gould92}
Gould. A., \& Loeb, A. 1992, \apj, 396, 104

\bibitem[Gould et al.(2010)]{gould10}
Gould. A., Subo, D., Gaudi, B. S., et al. 2010, \apj, 720, 1073

\bibitem[Gould et al.(2013)]{gould13}
Gould, A., Yee, J. C., Bond, I. A., et al. 2013, \apj, 763, 141

\bibitem[Han(1999)]{han99}
Han, C. 1999 \mnras, 309, 373

\bibitem[Han \& Gould(1995)]{han95}
Han, C., \& Gould, A. 1995, \apj, 447, 53

\bibitem[Han et al.(2005)]{han05}
Han, C., Gaudi, B. S., An. J. H., \& Gould, A. 2005, \apj, 618, 962 

\bibitem[Kayser et al.(1986)]{kayser86}
Kayser, R., Refsdal S., \& Stabell, R. 1986, \aap, 166, 36

\bibitem[Kim et al.(2010)]{kim10}
Kim, S.-L., Park, B.-G., Lee, C.-U., et al. 2010, \procspie, 7733, 77333F

\bibitem[Mao \& Paczy\'{n}ski(1991)]{mao91}
Mao, S., \& Paczy\'{n}ski, B. 1991, \apjl, 374, L37

\bibitem[Nataf et al.(2013)]{nataf13}
Nataf, D. M., Gould, A., Fouque, P., et al. 2013, \apj, 769, 88

\bibitem[Rhie et al.(2000)]{rhie00}
Rhie, S. H., Bennett, D. P., Becker, A. C., et al. 2000, \apj, 533, 378

\bibitem[Schneider \& Weiss(1987)]{schneider87}
Schneider, P., \& Weiss, A. 1987, \aap, 171, 49

\bibitem[Shvartzvald \& Maoz(2012)]{shvartzvald12} 
Shvartzvald, Y. \& Maoz, D. 2012, \mnras, 419, 3631

\bibitem[Sumi et al.(2003)]{sumi03}
Sumi, T., Abe, F., Bond, I. A., et al. 2003, \apj, 591, 204

\bibitem[Sumi et al.(2011)]{sumi11}
Sumi, T., Kamiya, K., Bennett, D. P., et al. 2011, \nat, 473, 349

\bibitem[Udalski(2003)]{udalski03}
Udalski, A. 2003, Acta Astron., 53, 291

\bibitem[Witt(1990)]{witt90}
Witt, H. J. 1990, \aap, 236, 311

\end{thebibliography}
\end{document}